\documentclass{iaus}
\usepackage{graphicx}
\usepackage{upmath}
\usepackage{amsbsy}
\usepackage{amsgen}

\newcommand{\OIII}{[O\,{\sc iii}]}

\title[JD 11.~~PNe in the Magellanic Clouds and Local Group] 
{Planetary Nebulae in the Magellanic Clouds and Local Group Galaxies}

\author[Warren A. Reid]   
{Warren A. Reid$^1~^2$
 }

\affiliation{$^1$Department of Physics and Astronomy, Macquarie University, Sydney, NSW 2109, Australia \\
$^2$Macquarie University Research Centre in Astronomy, Astrophysics \& Astrophotonics \\
email: {\tt warren.reid@mq.edu.au; war@aao.gov.au} \\[\affilskip]
}

\pubyear{2011}
\volume{283}  
\pagerange{1--8}
\jname{Planetary Nebulae an Eye to the Future}
\editors{A.C. Editor, B.D. Editor \& C.E. Editor, eds.}
\begin{document}

\maketitle

\begin{abstract}
The Magellanic Clouds are close enough to the Milky Way to provide an excellent environment in which to study extragalactic PNe. Most of these PNe are bright enough to be spectroscopically observed and spatially resolved. With the latest high resolution detectors on today's large telescopes it is even possible to directly observe a large number of central stars. Magellanic Cloud (MC) PNe provide several astrophysical benefits including low overall extinction and a good sample size covering a large range of dynamic evolutionary timescales while the known distances provide a direct estimation of luminosity and physical dimensions. Multi-wavelength surveys are revealing intriguing differences between MC and Galactic PNe.

Over the past 5 years there has been a substantial increase in the number of PNe discovered in the Large Magellanic Cloud (LMC) in particular. Deep surveys have allowed the faint end of the luminosity function to be investigated, finally providing a strong clue to its overall shape. In so doing, the surveys are approaching completeness, estimated at $\sim$80$\%$ in the LMC ($\sim$120 deg$^{2}$) and $\sim$65$\%$ in the Small Magellanic Cloud (SMC) ($\sim$20 deg$^{2}$).

The number of galaxies comprising the Local Group (LG) and its outskirts has been growing steadily over the past 5 years and now numbers 48. Most of the 7 newly discovered galaxies are dwarf spheroidal (dSph) in structure and range from 7.6 to 755 kpc from the Milky Way. Nonetheless, there are no published searches for PNe in any of these galaxies to date. Apart from the LMC and Milky Way, the number of PN discoveries has been very modest and only one additional LG galaxy has been surveyed for PNe over the past 5 years. This paper provides the number of Local Group PNe currently known and estimates each galaxy's total PN population.
\end{abstract}

\firstsection 

\section{The advantages of Local Group PNe outside the Milky Way}
Most astrophysical PN parameters, including ionized nebular mass
and the brightness and evolutionary states of their central stars,
depend on accurate distances (\cite[Ciardullo et al. 1999]{Ciardullo99}). This is
difficult in our own Galaxy due to inherent problems with variable
extinction and lack of central star homogeneity (\cite[Terzian, 1997]{Terzian97}). Since extra-galactic PNe are located outside the Milky Way they have the advantage of well established distances. The chemical composition of individual PNe within each galaxy can be directly compared in order to derive the intrinsic range in galactic chemical composition as it was a few Gyrs ago. Indeed, the oldest progenitors of today's observed PNe did not process elements such as S and Ar and to some extent also O and Ne. This makes them excellent probes for estimating the metallicity of the ISM at the time when the progenitors were agglomerating in the early life of their host galaxy. Elements produced and processed within the PN progenitor up until the end of the AGB including He, C, N and the $\textit{s}$-process elements are returned to the ISM, enriching material for the next generation of stars, allowing models of stellar evolution to be tested in a galactic context. Strong emission lines enable the measurement of electron temperatures and densities, which, along with emission line fluxes, provide estimates of central star temperatures and mass. Accurate radial velocities can also be measured using a number of emission lines.

These galaxies in turn represent a diverse mix of morphological types, metallicities, chemical evolution, star formation history and galactic evolution. Since all of these factors directly affect the observed PNe, including characteristics of their central stars and surrounding nebulae, a study of these objects becomes one of galactic archeology. Oxygen and neon abundances measured in PNe can be related to present day abundances in the ISM. By plotting the galactic location of observed PNe, models of past star formation and galactic evolution can be created.

\section{The advantages of Magellanic Cloud PNe}
PNe in the Magellanic Clouds are able to provide key data on the physics of stellar evolution,
mass loss (\cite[Iben 1995]{Iben95}), nucleosynthesis processes, abundance
gradients, ISM chemical enrichment and powerfully trace star-forming history (e.g. \cite[Maciel \& Costa 2003]{Maciel03}).
The well-determined 50~Kpc LMC distance (e.g. \cite[Keller \& Wood
2006]{Keller96}, \cite[Reid \& Parker 2010a]{Reid10a}), modest, 35~degree inclination angle and disk thickness (only
$\sim$500~pc, \cite[van der Marel \& Cioni 2001]{VanderMarel2001}), mean that LMC PNe are
effectively co-located. Since dimming of their light by intervening
gas and dust is generally low and uniform (e.g. \cite[Kaler \& Jacoby, 1990]{Kaler90}), absolute nebula luminosity and size can be estimated with minimal correction.

Although there are dwarf galaxies nearer to the Milky Way, few if any PNe have been found in them. The Magellanic Clouds provide single environments at known distances with luminosities up to 100 times that of the closer galaxies. Surveys in the Magellanic Clouds can not only identify candidate PNe but, with the benefit of known distance, provide initial emission-line luminosities, local environmental conditions and spatial details. With the combination of multi-wavelength data now available, including ground-based spectroscopy (eg. \cite[Reid \& Parker 2010a]{Reid10a}), HST imaging and UV spectroscopy (eg. \cite[Stanghellini et al 2005]{Stanghellini2005}), mid infrared data from the Spitzer Space Telescope survey of the LMC: Surveying the Agents of a Galaxy's Evolution (SAGE  \cite[Meixner et al (2006)]{Meixner2006}), deep $\textsl{YJK$_{s}$}$ photometry (eg. \cite[Cioni et al 2011]{Cioni2011}), and radio observations (eg. \cite[Filipovic et al 2009]{Filipovic2009}) the astrophysical tools are finally available to understand the effects of mass, temperature and metallicity during the PN formation and evolutionary cycles in each Magellanic Cloud.

The direct measurement and comparison of both nebulae and central stars across a wide evolutionary range allows the construction of extremely accurate nebula photoionization models. This task is more difficult to achieve in the Milky Way or more distant galaxies. Such powerful data contributes to our understanding of dredge-up effects, expansion velocities, changes according to dynamical age, nebula chemical contribution to the ISM, past stellar motions including those due to tidal forces and galactic evolution.

\section{PN populations}

Stellar population synthesis theory shows that the PN population size, when scaled to the available completeness limits, directly correlates with the visual magnitude of the galaxy (\cite[Magrini et al. 2003]{Magrini03}, \cite[Corradi \& Magrini 2006]{Corradi06}). This relation, however, is weakly dependent on the age, metallicity and the degree of star formation at the time the PN progenitors were forming. In addition, the PN specific luminosity rate (the proportionality coefficient: number of PNe L$^{-1}_{\odot bol}$) or ``$\alpha$ ratio" (\cite[Jacoby 1980]{Jacoby80}) may be strongly influenced by the morphological type of the host galaxy (\cite[Buzzoni, A., Arnaboldi, M. 2006]{Buzzoni06}).

When the theory of simple stellar populations (SSPs) (\cite[Renzini \& Buzzoni, 1986]{Renzini86} and \cite[Buzzoni, 1989]{Buzzoni89}) is applied to the Local Group, we have a basic reference frame allowing the evaluation of ``$\alpha$" for coeval and chemically homogeneous stellar conglomerates. In simple terms, the expected number of PN (N$_{PN}$) for a SSP of total luminosity (L$_{bol}$) is:
\begin{equation}
\textrm{N}_{PN} = {\ss}~\textrm{L}_{bol}~\tau_{PN}
\end{equation}

where $\ss$ is the so-called ``specific evolutionary flux" (which approximates to 2 $\times$ 10$^{-11}$ L$^{-1}_{\odot bol}$yr$^{-1}$; see \cite[Buzzoni (1989)]{Buzzoni89}) and $\tau_{PN}$ is the PN (emission-detectable) lifetime in years.

In reality, however, the stellar core mass evolution directly affects the PN lifetime since it depends on the central star temperature (T$_{eff}$ $\simeq$ 10$^{5}$ K) during the hot post AGB phase (see \cite[Renzini \& Buzzoni, 1986]{Renzini86}). Using models from \cite[Paczynski (1971)]{Pacz71} and \cite[Vassiliadis and Wood (1994)]{Vassiliadis94}, \cite[Buzzoni \& Arnaboldi (2006)]{Buzzoni06} have expressed this fuel in terms of H-equivalent solar mass. With an increase in the hot post AGB liftime with increasing dynamical ages and metallicities, the $\tau_{HPAGB}$ is somewhat equivalent to Fuel/$\textit{l}$$_{HPAGB}$, where $\textit{l}$$_{HPAGB}$ is the luminosity of the central star at the onset of the PN stage. The luminosity-specific PN density then becomes:

\begin{equation}
\alpha = \frac{\textrm{N}_{PN}}{\textrm{L}_{SSP}} = {\ss}\,\tau_{PN} = {\ss}\,\textrm{min}\{\tau_{HPAGB},\tau_{dyn}\}
\end{equation}

Despite physical age constraints that exist for the SSP, \cite[Buzzoni \& Arnaboldi (2006)]{Buzzoni06} have estimated a theoretical luminosity-specific PN density based on their equation (3.2) where
\begin{equation}
\alpha = 1 \textrm{PN} / 1.5 \times 10^{6} \textrm{L}_{\odot}.
\end{equation}

\begin{figure}[h]
\begin{center}
 \includegraphics[width=3.4in]{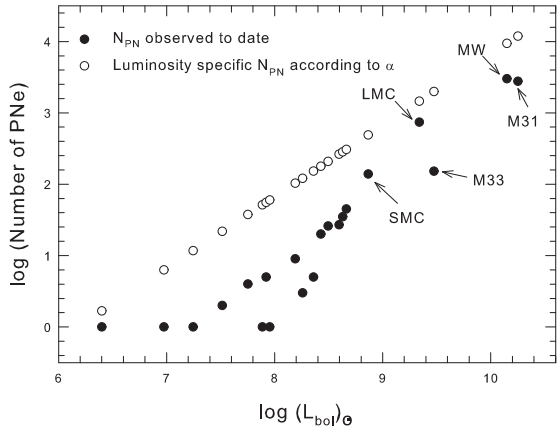}
 \caption{The number of PNe currently known in each galaxy as a function of the absolute estimated luminosity of their host galaxies. Filled circles represent the current census in each galaxy. Open circles represent the possible number of PNe predicted using $\alpha$, based on the theory of simple stellar populations.}
   \label{Figure 1}
\end{center}
\end{figure}

With a small bolometric correction applied to the L$_{v}$ of the galaxies, it is possible to use $\alpha$~to estimate the number of PNe that may exist in each galaxy. This bolometric correction does not amount to much more than -0.2 Mag. but appears to bring $\alpha$~closer to several independent population estimates eg. $\sim$1,000 PNe for the LMC (\cite[Reid \& Parker, 2006a,b]{Reid06a}), and 8,000 $\pm$ 1.500 for the Andromeda galaxy (\cite[Peimbert, 1990]{Peimbert90}). In order to compare the number of PNe currently uncovered and the likelihood of further discoveries, the known number of PN for each galaxy is plotted on the same scale and shown in Figure \ref{Figure 1}. The $\alpha$~estimate for each galaxy is shown by the open circles, which have a strong linear correlation. The main reason for real PNe deviating away from the same correlation, even at lower PN density levels, is the lack of a homogeneous completeness limit.
\begin{table}
  \begin{center}
  \caption{Known planetary nebulae populations in the local group and outlying galaxies in the years 2006 and 2011 showing the improvement. Local group galaxies newly identified/discovered in the past 5 years are indicated by an asterisk next to the name in column 1. }
  \label{tab1}
 {\scriptsize
  \begin{tabular}{lcrrrrll}
  \hline
Name	&	Type	&	Mv	&	Dist.	&	PNe	&	PNe	&	Ref (old)	&	Ref (new)	\\
	&		&		&	[kpc]	&	2006	&	2011	&	2006	&	2011	\\ [0.1cm]\hline\hline \\
M31	&	Sb	&	-21.2	&	785	&	2766	&	2766	&	Merrett 2006 	&		\\
Milky Way	&	Sbc	&	-20.9	&		&	~2400	&	~3000	&	Acker \textit{et al.} 1996	&	Parker \textit{et al.} 2006; \\
 & & & & & & & Miszalski  \textit{et al.} 2008	\\
M33	&	Sc	&	-18.9	&	795	&	152	&	152	&	Ciardullo \textit{et al.} 2004	&		\\
LMC	&	Ir	&	-18.5	&	50	&	277	&	740	&	Jacoby 2006 &	Reid 2006a,b, 2011$^{1}$ \\
SMC	&	Ir	&	-17.1	&	59	&	105	&	139	&	 Jacoby \textit{et al.} 2002	&	Jacoby 2006	\\
M32 (NGC221)	&	E2	&	-16.5	&	760	&	30	&	45	&	Ciardullo \textit{et al.} 1989 	&	Sarzi \textit{et al.} 2011 	\\
NGC205	&	Sph	&	-16.4	&	760	&	35	&	35	&	Corradi \textit{et al.} 2005	&		\\
IC10	&	Ir	&	-16.3	&	660	&	16	&	27	&	Magrini \textit{et al.} 2003	&	Kniazev, \textit{et al.} 2008 \\
NGC6822	&	dIr	&	-16.0	&	500	&	17	&	26	&	Leisy \textit{et al.}  2005	&	HM$^{2}$ \textit{et al.} 2009 	\\
NGC185	&	Sph	&	-15.6	&	660	&	5	&	5	&	Corradi \textit{et al.} 2005	&		\\
IC1613	&	dIr	&	-15.3	&	725	&	3	&	3	&	Magrini \textit{et al.} 2005 	&		\\
NGC147	&	Sph	&	-15.1	&	660	&	9	&	9	&	Corradi \textit{et al.} 2005	&		\\
WLM	&	dIr	&	-14.4	&	925	&	1	&	1	&	Magrini \textit{et al.} 2005	&		\\
Sagittarius	&	dSph/E7	&	-13.8	&	24	&	3	&	4	&	Zijlstra 1999	&	Zijlstra \textit{et al.} 2006	\\
Fornax (E351-G30)	&	dSph	&	-13.1	&	138	&	1	&	2	&	Danziger \textit{et al.} 1978 	&	Larsen 2008 	\\
Pegasus (DDO 216)	&	dIr	&	-12.3	&	760	&	1	&	1	&	Jacoby \textit{et al.} 1981	&		\\
Leo I (DDO 74)	&	dSph	&	-11.9	&	250	&		&		&		&		\\
Andromeda I	&	 IDsPH	&	-11.8	&	810	&		&		&		&		\\
Andromeda II	&	dSph	&	-11.8	&	700	&		&		&		&		\\
Leo A	&	dIr	&	-11.5	&	690	&	1	&	1	&	Magrini \textit{et al.} 2003	&		\\
DD 210	&	dIr	&	-11.3	&	1025	&		&		&		&		\\
Sag DIGD	&	dIr	&	-10.7	&	1300	&		&		&		&		\\
Pegasus II	&	dSph	&	-10.6	&	830	&		&		&		&		\\
Pisces (LGS3)	&	dIr	&	-10.4	&	810	&		&		&		&		\\
Andromeda V	&	dSph	&	-10.2	&	810	&		&		&		&		\\
Andromeda III	&	dSph	&	-10.2	&	760	&		&		&		&		\\
Leo II (Leo B)	&	dSph	&	-10.1	&	210	&		&		&		&		\\
Cetus*	&	dSph	&	-9.9	&	755	&		&		&		&		\\
Phoenix	&	dSph	&	-9.8	&	395	&		&	1	&		&	Saviane \textit{et al.} 2009 	\\
Sculptor (E351-G30)	&	dSph	&	-9.8	&	87	&		&		&		&		\\
Cassiopeia (An VII)	&	dSph	&	-9.5	&	690	&		&		&		&		\\
Tucana	&	dSph	&	-9.6	&	870	&		&		&		&		\\
Sextans	&	dSph	&	-9.5	&	86	&		&		&		&		\\
Carina (E206-G220)	&	dSph	&	-9.4	&	100	&		&		&		&		\\
Draco (DDO 208)	&	dSph	&	-8.6	&	79	&		&		&		&		\\
Ursa Minor	&	dSph	&	-8.5	&	63	&		&		&		&		\\
Canes Venatici I*	&	dSph	&	-7.8	&	220	&		&		&		&		\\
Leo T*            &   dSph     &    -7.1    &    420  &     &       &      &    \\
Ursa Major*	&	dSph	&	-6.7	&	100	&		&		&		&		\\
Canis Major Dwarf* 	&	Irr	&		&	7.6	&		&		&		&		\\
Canes Venatici II*	&	dSph	&	-5.8	&	150	&		&		&		&		\\
Bootes*	&	dSph	&	-5.8	&	60	&		&		&		&		\\
Ursa Major II*	&	dSph	&	-3.8	&	30	&		&		&		&		\\
	&		&		&		&		&		&		&		\\
\textit{LG outskirts}	&		&		&		&		&		&		&		\\
GR8	&	dSph	&	-11.8	&	2200	&	0	&		&	Magrini \textit{et al.} 2005 	&		\\
Antlia	&	dSph	&	-15.8	&	1330	&		&		&		&		\\
NGC3109	&	dIr	&	-15.8	&	1330	&	18	&	20	&	Corradi \textit{et al.} 2006 	&	Pe\~{n}a \textit{et al.} 2007	\\
Sextans B	&	dIr	&	-14.3	&	1600	&	5	&	5	&	Magrini \textit{et al.} 2000	&		\\
Sextans A	&	dIr	&	-14.2	&	1320	&	1	&	1	&	Magrini \textit{et al.} 2003	&		\\
EGB0427+63	&	sIr	&	-10.9	&	2200	&		&		&		&		\\ \hline
  \end{tabular}
  }
 \end{center}
\vspace{1mm}
 \scriptsize{
 {\it Notes:}\\
  $^1$This presentation.
  $^2$Hern$\acute{a}$ndez-Mart$\acute{i}$nez et al. (2009).
  }
\end{table}
The number of spectroscopically confirmed and candidate PNe currently known in the Local Group galaxies is shown in column 6 of Table~\ref{tab1} (2011). Column 5 gives the number of PNe known at the time of the previous PN conference (IAUS234 in 2006). It can be seen that the largest increase in the number of candidate PNe has occurred in the LMC and in our own Galaxy. Other Local Group galaxies record only moderate increases, but the numbers expected to be found will also be much lower (see Figure~\ref{Figure 1}) due to their lower overall stellar populations. Column 7 provides a reference to the 2006 census whereas column 7 provides a reference to the most up to date and extensive census work.

\section{Multi-wavelength PN surveys and research in the Magellanic Clouds over the past 5 years}

Over the past 5 years there have been 16 published papers that specifically analysed PNe in both the LMC and SMC, 11 of which employed space-based observations. In addition there were individually 8 LMC and 6 SMC PN papers published, 8 of which used space-based data. Overall, 63\% of the PNe studies in the MCs over the past 5 years employed space-based observations using Spitzer Space Telescope (NIR), HST (optical and UV), FUSE and XMM-Newton satellite (x-ray). The past 5 years has importantly seen major improvements to radio telescope detectors, leading to the mapping of radio emission from MC PNe. These detections have already become the subject of 4 papers.
\subsection{Optical observations}
The largest and most productive survey for PNe in the LMC over the past 5 years has been that conducted by Reid \& Parker (2006a,b). This UK Schmidt Telescope survey of the central 25deg$^{2}$ main bar region used the best 12 best deep H$\alpha$ and 6 short red images, obtained over a 3 year period, which were then scanned and combined using SuperCOSMOS. Candidate identification was carried out by assigning a separate colour to each digital image and then overlaying the two. This method is somewhat different to the usual methods of on-band/off-band \OIII~imaging and wide-field slitless spectroscopy such as objective prism imaging. The process is fully described in Reid \& Parker (2006a). The resulting candidates were spectroscopically followed up using a variety of telescopes including 2dF on the AAT, and FLAMES on the VLT leading to new LMC kinematic studies (\cite[Reid \& Parker 2006b]{Reid2006b}), a new PN luminosity function (\cite[Reid \& Parker 2010a]{Reid2010a}) and a new improved method for determining excitation class (\cite[Reid \& Parker 2010b]{Reid2010b}) .

\begin{figure}[h]
\begin{center}
 \includegraphics[width=5.0in]{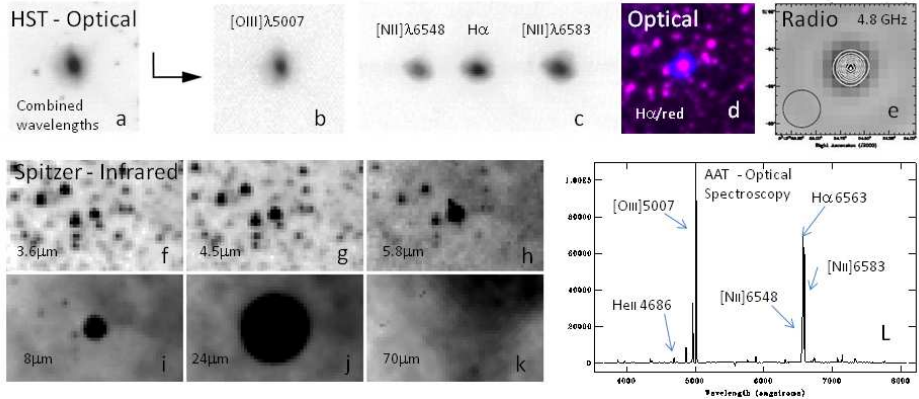}
 \caption{Multi-wavelength data for LMC PN SMP47. a-c = HST data, d = UK Schmidt Telescope deep ground-based imaging, e = ATCA + Parkes with 8.64 GHz overlaid contours, f-k = Spitzer Space Telescope infrared data, L = optical spectroscopy from 2dF on the AAT.}
   \label{Figure 2}
\end{center}
\end{figure}

Over 2,000 candidate objects of different types were spectroscopically confirmed, including PNe in various stages of evolution which were released with a quality/certainty flag. Continuing multi-wavelength observations (see Figure \ref{Figure 2}) have secured classifications, reducing the number of candidates in the lowest quality (possible PN) and uncovering 93 new PNe in the outer LMC (see \cite[Reid \& Parker 2011]{RP11}).
\subsection{Abundances}
Using the empirical method for abundance determination which applies ionisation correction factors (ICFs), \cite[Leisy \& Dennefeld (2006)]{Leisy2006} found a slight anti-correlation between N and O for MC PNe. This relation which is more pronounced in the SMC is probably due to higher efficiency of the surface enrichment through dredge-ups which is expected to occur at lower metallicities. With Type\,I PNe defined as He/H $>$ 0.10 and N$^{+}$/O$^{+}$ $>$ 0.25, at least 21 PNe from both clouds were found to agree with the N/O requirement but low in He (He/H $<$0.10). These PNe, named as Type\,I$\textit{i}$, were also found by \cite[Reid 2007]{Reid07} in an independent study which used the same method for abundance determination. By plotting N/O versus O, \cite[Leisy \& Dennefeld (2006)]{Leisy2006} found that the destruction of O in high mass stars is greater in the LMC and SMC than in our Galaxy. A similar plot for PNe in the SMC by \cite[Shaw \etal\ 2010]{Shaw10} gives the impression that there is no substantial decrease in N/O with increasing O. These same authors did not find Type\,I PNe in their observed sample but made mention that SMP PN MG8, a non Type\,I PN, had an estimated central star mass of 0.88M$_{\odot}$ implying a main sequence mass $>$ 3.5M$_{\odot}$, which is close to the minimum mass where SMC metallicites would allow hot bottom burning to convert C to N.
\subsection{Infrared observations}
The growing use of Spitzer SAGE data (\cite[Meixner \etal\ 2006]{Meixner2006}) is enabling far more detailed analysis of compact and point-like objects such as PNe in the Magellanic Clouds.
With the increasing use of this data comes a new understanding of the evolutionary properties of PNe and the effect that the
metallicity of the host system imposes on the observed spectra. Using the brightest 233 LMC PNe, \cite[Hora et al. (2008)]{Hora08}
found low detection rates, for example, only 119 sources in all 4 IRAC bands, 185 in 2 bands and 28 in only 1 IRAC band. In MIPS they found 161, 18 and 0 matching the 24, 70 and 160$\mu$m bands respectively. The larger number of detections at 24$\mu$m probably represents the high sensitivity of this MIPS array and the peak of the SED for many PNe which correspond to the temperature of $\sim$100K for dust inside the ionised zone, trapped by Ly$\alpha$ photons. Low detection rates in other bands may indicate that dust and PAHs are evolving according to the metallicity, temperature and UV luminosity of the nebula.
Compared to Galactic PNe adjusted to the LMC distance, the LMC sample has similar 3.6$\mu$m to 8$\mu$m line strengths but the sensitivity cutoff for SAGE passes roughly through the middle of the Galactic distribution indicating that a large sample of LMC PNe (highly evolved and faint ones) are not likely to be detected. Crowding and completeness limits are likely to contribute to the non-detections, however using those that could be detected \cite[Hora et al. (2008)]{Hora08} began to find curious differences between Galactic and LMC PNe.

Areas of IRAC colour-colour plots which are normally associated with galaxies, AGNs, and young stellar objects (YSOs) due to star-forming regions in the LMC, are also occupied by a large proportion of LMC PNe. In fact, LMC PNe occupy and extend all the way to the region populated by stars at 0-0 colour points. Such PNe (eg. J21 and J22) are dominated by their central stars indicating that they are highly evolved PNe in which the gas is recombining, thereby eliminating other characteristic MIR processes. A MIR survey, isolated from other spectral regimes, may well mistake these objects for YSOs, and this is likely to have already occurred. Worse still, at least one attempt to classify or re-classify MC PNe using J, Y and K$_{s}$ colours alone has lead to confusion and errors. It is clearly risky to attempt to identify MC PNe when IR detections are below SAGE limits and optical spectra are not obtained.

Strong PAH features are typical of young PNe. In more evolved PNe the PAHs are dissolving and H recombination lines dominate the near-IR emission (\cite[Hora et al. 2008]{Hora08}).

Colour comparisons to 8$\mu$m emission show peculiarities of LMC PNe that may be linked to low metallicity and/or evolution of the nebulae. Since the 5.8$\mu$m and 8$\mu$m bands both include PAH emission features, the strengths are well correlated but when comparing 4.5$\mu$m to 8$\mu$m for the same objects the greater spread may be due in part to a lack of PAHs and much fainter emission from warm dust at 4.5$\mu$m. Going further to the blue and comparing 3.6$\mu$m to 8$\mu$m colours separates PNe such as SMP\,11 which occupy an area known for extreme AGB stars. While the spectrum of SMP\,11 clearly indicates a PN, its MIR colours suggest a post-AGB object (\cite[Bernard-Salas et al. 2006]{Salas06}). Comparing 8$\mu$m versus $\textsl{J}$-8$\mu$m, LMC PNe are appear brighter and redder than the Galactic PNe, but this may include a sampling effect since the faintest LMC PNe were not included.

Great care needs to be taken when interpreting the spectral energy distribution (SEDs) for MC PNe as the shape of the SED is strongly influenced by the mass and evolutionary state of the PN. When MIR bands for bright, young MC PNe with detectable emission at 70$\mu$m are plotted, the SEDs all move in a similar line. This is not the case for more evolved objects where \cite[Hora et al. (2008)]{Hora08} show a much more diverse range of SEDs, many of which decrease towards 8$\mu$m and even decrease from 8$\mu$m to 24$\mu$m. Similar features were found by \cite[Stanghellini et al (2007)]{Stanghellini07}, who, with the advantage of complete MIR spectra found that many LMC PNe show an absence of grain emission and can be categorised as ``featureless". PNe with a low dust continuum and the absence of features have not been observed in the Galaxy. PNe with a prominent silicon carbide feature at 11$\mu$m are those with carbon-rich dust, categorised as type CRD. PNe with crystalline silicate features at 23.5, 27.8 and 33.5$\mu$m have oxygen-rich dust and may categorised as type ORD.

With these 3 categories identified, the IR luminosity versus carbon abundance shows that the carbon-rich PNe are also among the most luminous. Small samples indicate they are either round or bipolar core in morphology. Oxygen-rich PNe are also bright but may be as much as 1 log C/H lower in carbon abundance than carbon-rich PNe. They appear to be bipolar or quadrupolar in morphology. Featureless PNe extend from the mid-luminosity region occupied by carbon-rich PNe but then show a steady decline in both luminosity and carbon abundance. They appear to be either round or bipolar core when carbon-rich but mainly bipolar core when carbon poor. The dust temperature also decreases with decreasing IR luminosity where featureless PNe dominate the lower end on both scales. These findings show that LMC PNe show extreme changes in their MIR and JHK features with evolution. Future work with larger samples will enable clear definitions to be formulated.

\end{document}